\documentclass[amsmath,amssymb,prl,hyperlink,twocolumn]{revtex4}

	\usepackage{graphicx}
	\usepackage{soul}
	\usepackage[colorlinks=true,citecolor=blue,linkcolor=magenta]{hyperref}
	\usepackage[usenames]{color}
	\usepackage{amsfonts}
	\usepackage{color}
	\usepackage{multirow}
	\usepackage{float}
\usepackage{times}
\usepackage{graphicx}
\usepackage{gensymb}
\usepackage{braket}
\usepackage{amsmath}
\usepackage{textcomp}
\usepackage{color}

\usepackage[colorlinks=true,citecolor=blue,linkcolor=magenta]{hyperref}

\begin{document}

\title{Solving Systems of Linear Equations with a Superconducting Quantum Processor}

\author{Yarui Zheng,$^{1,3*}$ Chao Song,$^{2,3}$\footnote{Y. Z. and C. S. contributed equally to this work.} Ming-Cheng Chen,$^{3,4,5}$ Benxiang Xia,$^{3,4,5}$ Wuxin Liu,$^{2}$ Qiujiang Guo,$^{2}$ Libo Zhang,$^{2}$ Da Xu,$^{2}$ Hui Deng,$^{1}$ Keqiang Huang,$^{1}$ Yulin Wu,$^{1}$ Zhiguang Yan,$^{1}$ Dongning Zheng,$^{1}$ Li Lu,$^{1}$ Jian-Wei Pan,$^{3,4,5}$ H. Wang,$^{2,3}$\footnote{Electronic mail: hhwang@zju.edu.cn} Chao-Yang Lu,$^{3,4,5}$\footnote{Electronic mail: cylu@ustc.edu.cn} and Xiaobo Zhu$^{1,3,4,5}$\footnote{Electronic mail: xbzhu16@ustc.edu.cn} \vspace{0.2cm}}

%\thanks{Y. Z. and C. S. contributed equally to this work.}
\affiliation{$^1$ Institute of Physics, Chinese Academy of Sciences, Beijing 100190, China}
\affiliation{$^2$ Department of Physics, Zhejiang University, Hangzhou, Zhejiang 310027, China}
\affiliation{$^3$ CAS Centre for Excellence and Synergetic Innovation Centre in Quantum Information and Quantum Physics, University of Science and Technology of China, Hefei, Anhui 230026, China}
\affiliation{$^4$ Hefei National Laboratory for Physical Sciences at Microscale and Department of Modern Physics, University of Science and Technology of China, Hefei, Anhui 230026, China}
\affiliation{$^5$ CAS-Alibaba Quantum Computing Laboratory, Shanghai, 201315, China}

\date{\today}

\begin{abstract}
Superconducting quantum circuits are promising candidate for building scalable quantum computers.
Here, we use a four-qubit superconducting quantum processor to solve a two-dimensional system
of linear equations based on a quantum algorithm proposed by Harrow, Hassidim, and Lloyd [Phys. Rev. Lett. \textbf{103}, 150502 (2009)],
which promises an exponential speedup over classical algorithms under certain circumstances.
We benchmark the solver with quantum inputs and outputs, and characterize it
by non-trace-preserving quantum process tomography, which yields a process fidelity
of $0.837\pm0.006$. Our results highlight the potential of superconducting quantum circuits
for applications in solving large-scale linear systems, a ubiquitous task in science and engineering.
\end{abstract}

%\keywords{linear equations, quantum computer, quantum algorithm, superconducting qubit}

\maketitle


Linear system lies at the heart of many areas of science and engineering.
To solve a system of linear equations with $N$ variables, the best known
classical algorithm requires a time of $O(N)$. Harrow, Hassidim, and Lloyd
(HHL) \cite{PRL.103.150502} showed that in principle quantum computers
can solve linear systems exponentially faster
by calculating the expectation value of an operator associated with the solution,
which may lead to many practical applications of quantum computation other than those previously known \cite{Grover1996,Deutsch&Jozsa1992,P.Shor1994}.
For an $s$-sparse system matrix of size $N{\times}N$ and condition number $\kappa$,
the HHL algorithm can reach a desired computational accuracy $\epsilon$ within
a running time of $O(log(N)s^2\kappa^2/\epsilon)$ under certain circumstances~\cite{Aaronson2015},
comparing to the best known classical algorithm of $O(Ns\kappa/log(\epsilon))$.
Such an exponential speedup
promises widespread applications that address large-scale systems.
Indeed, several applications based on the HHL algorithm, such as
data processing \cite{PhysRevLett.109.050505}, numerical calculation \cite{PhysRevA.93.032324}
and artificial intelligence \cite{arXiv1307.0411, PhysRevLett.113.130503},
have been proposed in recent years.

Compiled version of the HHL algorithm was previously only demonstrated
with parametric down-converted single photons \cite{PRL.110.230501, srep06115}
and liquid nuclear magnetic resonance \cite{PhysRevA.89.022313},
both of which are considered not easily scalable to large number of qubits.
For example, the optical demonstration was limited by the probabilistic photon generation
and two-photon gate operation. For a deterministic, and more scalable implementation,
here we turn to a solid-state system, i.e., a superconducting quantum circuit in this experiment,
which has attracted significant attentions due to a number of merits, including
the much-improved coherence~\cite{Paik2011,Barends2013,Rigetti2012},
the excellent scalability~\cite{Kelly2015,Corcoles2015,Chen2014},
and the remarkable high-fidelity and fast control~\cite{Barends2014,Sheldon2016,Riste2013}.
%\cite{ Science.285.5430, Nature398786a, PhysRevLett.89.117901, PhysRevA.76.042319}
In addition, compiled versions of various quantum algorithms
such as the Deutsch-Jozsa algorithm~\cite{DiCarlo2009,Yamamoto2010}, the Grover's algorithm~\cite{DiCarlo2009},
and the Shor's algorithm~\cite{Lucero2012} have been successfully tested on this solid-state platform on a small scale.

In this Letter, we demonstrate a nontrivial instance of a quantum linear solver,
based on the HHL algorithm for a $2\times2$ system, with a superconducting circuit
consisting of four Xmon qubits which are variant of the transmon qubits~\cite{Koch2007}.
We test the solver with 18 distinct quantum-state inputs that uniformly distribute on the Bloch sphere,
from which the non-trace-preserving quantum process tomography (QPT) can be reliably determined \cite{CHUANG&Nielsen1997}.
For various quantum inputs,
our quantum solver can return the desired solutions with reasonably high precision,
yielding an averaged QPT fidelity of $0.837\pm0.006$.
As such our experiment represents the first demonstration of the quantum algorithm for solving systems of linear equations on a solid-state platform.

The device was fabricated on the sapphire substrate in three steps following
the procedure outlined previously~\cite{Barends2013}:
(1) deposit the aluminum film on the degassed substrate;
(2) define circuit wirings using wet-etch;
(3) double-angle evaporate the Al/AlO$_\textrm{x}$/Al Josephson junctions.
Figure~1 shows the optical micrograph of the device,
with the four Xmon qubits labeled from $Q_1$ to $Q_4$.
Each qubit has its own frequency-control Z line, for rotations of the qubit state
around Z axis on the Bloch sphere.
$Q_1$ and $Q_2$ ($Q_3$ and $Q_4$) share the microwave XY line
on the left (right) that is closer to $Q_1$ ($Q_4$),
for single-qubit rotations around X and Y axes.
The microwave pulses transmitted through each XY line have two-frequency components,
and the drive strength to $Q_1$ ($Q_4$) is more than that to $Q_2$ ($Q_3$) by a factor of 11.7 (6.7) as experimentally calibrated.
Each qubit dispersively couples to its own readout resonator,
and all readout resonators couple to a common transmission line,
enabling simultaneous single-shot quantum nondemolition measurement
on multiple qubits using frequency-domain multiplexing.
The signal-to-noise ratio is further improved by a quantum limited
parametric amplifier, similar to that described previously \cite{APL.104.263513}.

The circuit Hamiltonian under the rotating wave approximation is
\begin{equation}
\label{Ham}
H = -\sum_{j=1}^4{\omega_j(t)\sigma_j^z/2}+\sum_{j=1}^3{g_{j,j+1}\left(\sigma_j^+\sigma_{j+1}^-+\sigma_j^-\sigma_{j+1}^+\right)},
\end{equation}
where $\omega_j(t)$ is the resonant frequency of the $j$-th qubit that can be tuned over time,
$\sigma_j^z$ is the Pauli operator, $\sigma_j^{\pm}$ are the raising and lowering operators,
and $g_{j,j+1}$ is the nearest-neighbor coupling strength.
The coupling strengths of $g_{j,j+1}/2\pi$, for $j = 1$ to 3, are measured to be around 13.0, 9.8, and 14.1~MHz, respectively.
These Xmon qubits typically have a maximum frequency around 5.1 GHz and an anharmonicity around 250~MHz.
In this experiment, qubit idle frequencies $\omega_j(t=0)/2\pi$, for $j = 1$ to 4, are arranged in a zigzag pattern at
5.073, 4.074, 4.948, and 4.547 GHz, respectively, which ensures that
the nearest-neighbor and next nearest-neighbor couplings
along the qubit chain are effectively turned off when idling.
At the above-listed frequencies, the qubit lifetimes $T_1$s are measured to be around 15.9, 7.4, 7.8, and 14.1~${\mu}$s,
and the Gaussian dephasing times~\cite{Sank2012} $T_2^\ast$s are around 8.7, 2.3, 5.2, and 3.4~${\mu}$s, respectively
(see \cite{Supplement} for more information on the device).

\begin{figure}[t]
\includegraphics[width=\linewidth]{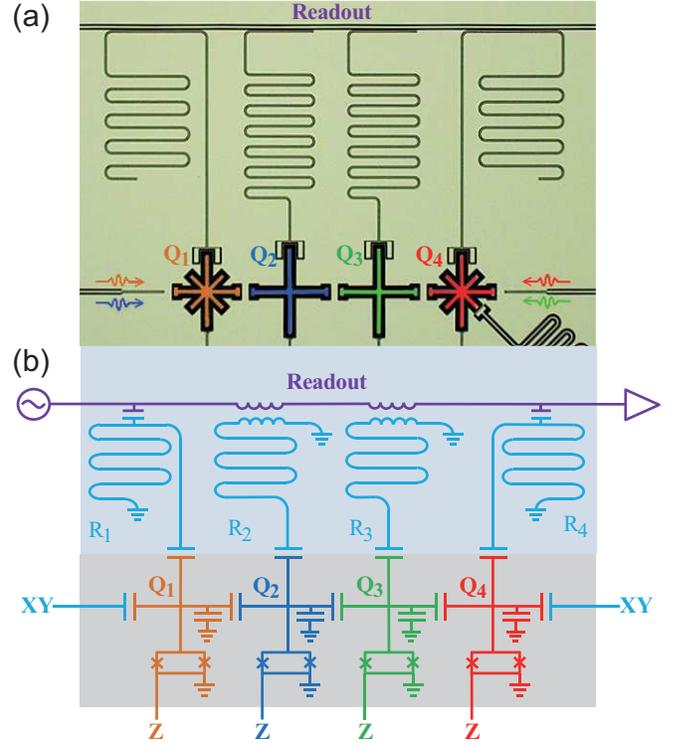}
\caption{(a) False color photomicrograph and (b) simplified circuit schematic of the superconducting quantum circuit
for solving $2 \times 2$ linear equations. Shown are the four Xmon qubits, marked from $Q_1$ to $Q_4$,
and their corresponding readout resonators, marked from $R_1$ to $R_4$.
\label{figure_sample}}
\end{figure}

The HHL algorithm aims to solve a system of linear equations $A\vec{x}=\vec{b}$ for $\vec{x}$,
given the $N{\times}N$ Hermitian matrix $A$ and the input vector $\vec{b}$.
The process involves three subsets of qubits: a single ancilla qubit, a register
of $k$ qubits used to store the eigenvalues of $A$ to a binary precision of
$k$ bits, and a memory of $O\left(\log(N)\right)$ qubits used to load $\vec{b}$ and also store the output $\vec{x}$.
For simplicity we assume that $\vec{b}$ is a unit vector,
whose entries \{$b_i$\} can be encoded in the memory
formatted to a quantum state $\ket{b} = \sum{b_i\ket{i}}$, where $\ket{i}$ denotes the
computational basis of the $O\left(\log(N)\right)$ qubits.
Next is the core of the HHL algorithm responsible for the exponential speedup:
with carefully designed quantum logic gates including mapping the Hermitian matrix $A$ to the system Hamiltonian,
the quantum state $\ket{x}$ representing the desired solution $\vec{x}$
can be synthesized in the memory conditional upon the state of the ancilla qubit.
Afterward one can either map the quantum state $\ket{x}$ to recover all entries of the vector $\vec{x}$,
or, more efficiently, perform the quantum measurement corresponding to
an operator $M$, that one is interest in, to extract its expectation value $\langle x|M|x\rangle$.

\begin{figure*}[t]
\includegraphics[width=0.94\linewidth]{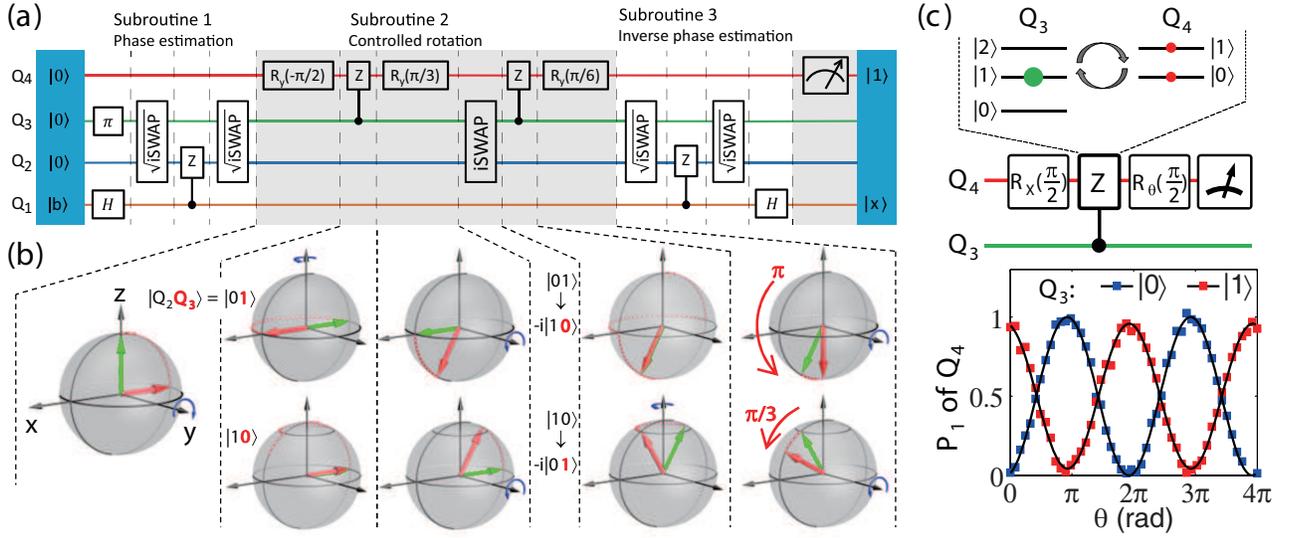}
\caption{(a) Compiled quantum circuits for solving $2 \times 2$ linear equations with four qubits.
There are three subroutines and more than 15 gates as indicated.
(b) The Bloch-sphere illustration of the controlled rotation subroutine for $C=1$, where
the two rotation angles of $\pi$ and $\pi/3$ are achieved for eigenvalues $\lambda_1 = 1$ ($\ket{01}$) and $\lambda_2 = 2$ ($\ket{10}$), respectively.
(c) The $Q_3$-$Q_4$ CZ gate. Top: qubit energy level arrangement showing that
the $\ket{0} \leftrightarrow \ket{1}$ transition of the target $Q_4$ is on resonance with
the $\ket{1} \leftrightarrow \ket{2}$ transition of the control $Q_3$.
Only when $Q_3$ is in $\ket{1}$, the state in $Q_4$ will make a full cycle and gain an additional phase of $\pi$.
Middle: the quantum circuits for calibrating the CZ gate sandwiched in between two $\pi/2$ rotations, where the second $\pi/2$
rotation axis has an angle $\theta$ from X axis in XY plane of the Bloch sphere.
Bottom: the calibrated Ramsey interference curves of $Q_4$ when $Q_3$ is in $\ket{0}$ (blue) and $\ket{1}$ (red), which differ by a phase of $\pi$.
No single-qubit phase gates were used during this measurement to cancel the dynamical phase due to the change of qubit frequency.
\label{figure_quantumcircuits}}
\end{figure*}

With the system initialized in the state $\ket{0}_\textrm{a}\ket{0}_\textrm{r}\ket{b}_\textrm{m}$, where the subscripts
a, r, and m index, respectively, the subsets of qubits in the ancilla, the register, and the memory (here
and below we keep the subscripts in wavefunctions only when the states of two or three subsets are quoted simultaneously),
a general description of the HHL core is as follows:
(1) with quantum phase estimation \cite{PRL.103.150502,NielsenChuangbook} using the controlled unitary transformations
in the form $e^{-iAt}$ for a variable time $t$,
decompose $\ket{b}$ in the eigenbasis of $A$, i.e., $\ket{b} = \sum_j{\beta_j\ket{u_j}}$, and map the
corresponding eigenvalues $\lambda_j$ into the register in a binary form to transform
the system to $\sum_j{\beta_j\ket{0}_\textrm{a}\ket{\lambda_j}_\textrm{r}\ket{u_j}_\textrm{m}}$;
(2) perform controlled rotation $R(\lambda^{-1})$ on the ancilla
according to $\lambda_j$ stored in the register,
which transforms the system to
\begin{equation}
\label{CR}
\sum_{j}{\beta_j\left(\sqrt{1-\frac{C^2}{\lambda_j^2}}\ket{0}_\textrm{a}+\frac{C}{\lambda_j}\ket{1}_\textrm{a}\right)\ket{\lambda_j}_\textrm{r}\ket{u_j}_\textrm{m}},
\end{equation} where $C$ ($\leq 1$) is a constant that can be selected
as any real number to make the controlled rotation physical \cite{PRL.103.150502};
(3) reverse the procedure in (1) to disentangle and clear the register, and
the system state evolves to
\begin{equation}
\sum_{j}{\beta_j\left(\sqrt{1-\frac{C^2}{\lambda_j^2}}\ket{0}_\textrm{a}+\frac{C}{\lambda_j}\ket{1}_\textrm{a}\right)\ket{0}_\textrm{r}\ket{u_j}_\textrm{m}}.
\end{equation}
A postselection of the $\ket{1}$-state outcome of the ancilla
will yield the desired output in the memory $\ket{x} \sim \sum_j{C(\beta_j/\lambda_j)\ket{u_j}}$,
with a success probability of $\sum_j{\left(C\beta_j/\lambda_j\right)^2}$.

As argued elsewhere~\cite{Aaronson2015}, conversions between the
classical vectors and their quantum counterparts, i.e., $\vec{b} \leftrightarrow \ket{b}$
and $\vec{x} \leftrightarrow \ket{x}$, take extra time and
may eventually kill the exponential speedup gained during execution of the HHL core.
Nevertheless, the HHL algorithm provides a general template and represents
a real advance in the theory of quantum algorithms. Our immediate goal
in this experiment is to implement a \emph{purely} quantum version of the HHL algorithm,
i.e., we aim to test the above-mentioned HHL core with quantum inputs and outputs.

In our demonstration, the four-qubit solver is set to run a nontrivial instance,
where the system matrix $A$ is chosen as
%\begin{equation}
$A = \left(\begin{smallmatrix} 1.5&0.5\\0.5&1.5\protect \end{smallmatrix}\right)$.
%\end{equation}
Eigenvectors of $A$ formatted to quantum are
$\ket{u_{1}} = \left(\ket{0}-\ket{1}\right)/\sqrt{2}$ and $\ket{u_{2}} = \left(\ket{0}+\ket{1}\right)/\sqrt{2}$
with eigenvalues of $\lambda_{1} = 1$ and $\lambda_{2} = 2$, respectively.
Accordingly, our four qubits are distributed into three subsets: the ancilla ($Q_4$), the register ($Q_2$$Q_3$), and the memory ($Q_1$).
The binary representations of ``$Q_2Q_3$'', ``$01$'' and ``$10$'', record eigenvalues $\lambda_1$ and $\lambda_2$, respectively.
The input $\ket{b}$ is prepared in $Q_1$ by single-qubit rotations.

\begin{figure}[th]
\includegraphics[width=0.96\linewidth]{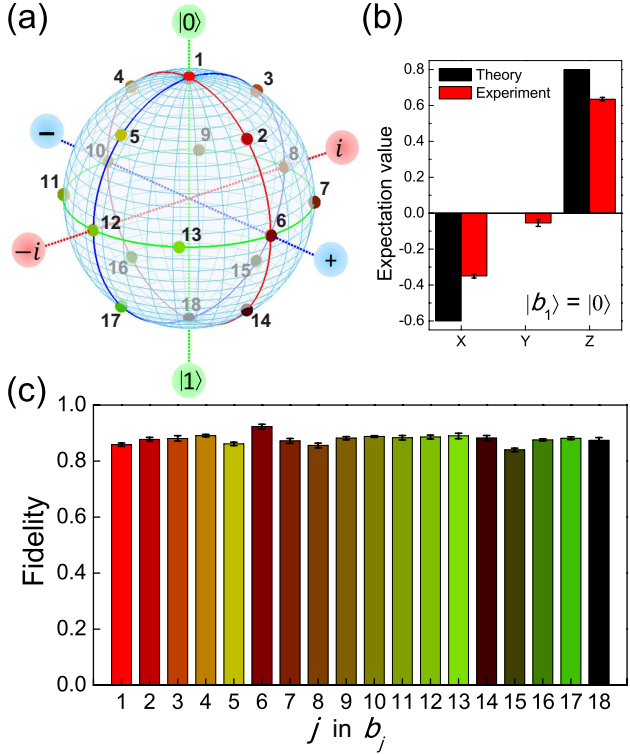}
\caption{(a) 18 input quantum states indexed by the number $j$ on the Bloch sphere that are
used to test the quantum linear solver.
(b) Expectation values of the three operators, where \{$X$, $Y$, $Z$\} are the
Pauli operators \{$\sigma_x$, $\sigma_y$, $\sigma_z$\}, for the output state $|x\rangle$ when the input state $\ket{b_1} = \ket{0}$.
(c) Fidelity values of the output states corresponding to the 18 input states $\ket{b_j}$ that are labeled
on the Bloch sphere in \textbf{a}.
Statistical errors are shown by error bars,
defined as $\pm 1$ s.d., using the repeated sets of the QST measurement.
\label{figure_resultfidelity}}
\end{figure}

The pre-chosen $2\times2$ Hermitian matrix $A$ allows us to optimize the circuit
that consists of three subroutines as shown in Fig.~2(a),
where all two-qubit gates fit to our system Hamiltonian (Eq.~\ref{Ham}).
\textbf{Subroutine 1:} the phase estimation subroutine is pre-compiled with a controlled-phase (CZ) gate, two $\sqrt{\textrm{iSWAP}}$
gates, and two single-qubit gates acting on the memory ($Q_1$) and the register ($Q_2Q_3$),
which can be described as follows:
first prepare
$\ket{b} = \beta_1 \ket{u_1} + \beta_2 \ket{u_2}$
in $Q_1$,
following which the Hadamard gate transforms $Q_1$ to
$\ket{b} = \beta_1 \ket{1} + \beta_2 \ket{0}$;
meanwhile a $\pi$ rotation on $Q_3$ yields
$\ket{01}$
in the $\ket{Q_2Q_3}$ register;
next the sandwiched $\sqrt{\textrm{iSWAP}}$-CZ-$\sqrt{\textrm{iSWAP}}$ gate combo
fulfills a controlled-iSWAP gate, which swaps the states
between $Q_2$ and $Q_3$ up to a phase factor of $-i$ only if $Q_1$ is in $\ket{0}$.
At the end of this subroutine, the state of $\ket{Q_2Q_3}_\textrm{r}\ket{Q_1}_\textrm{m}$
goes to $\beta_1 \ket{01}_\textrm{r} \ket{1}_\textrm{m} - i\beta_2 \ket{10}_\textrm{r} \ket{0}_\textrm{m}$,
which correlates the binary representations of the eigenvalues in the register
(ignoring the phase factor as we omit an insignificant single-qubit phase gate here)
with the eigenstates in the memory, since
$\ket{1}$ and $\ket{0}$ correspond to
$\ket{u_1}$ and $\ket{u_2}$,
respectively, up to a Hadamard gate.
\textbf{Subroutine 2:} the controlled rotation $R(\lambda^{-1})$ subroutine acts on the ancilla ($Q_4$)
depending on the state of the register $\ket{Q_2Q_3}$, whose effect can be best visualized using the Bloch sphere
representation as illustrated Fig.~2(b). Here we choose $C = 1$, and according to Eq.~\ref{CR}
the rotation angles around Y axis for $Q_4$ should be $\pi$ for $\lambda_1 = 1$ and $\pi/3$ for $\lambda_2 = 2$.
\textbf{Subroutine 3:} this one is exactly the reverse of Subroutine 1, in which
the register $\ket{Q_2Q_3}$ is disentangled and cleared, and
$\ket{1}$ and $\ket{0}$
are transformed back to the eigenstates
$\ket{u_1}$ and $\ket{u_2}$,
respectively,
with the final Hadamard gate on $Q_1$.

The compiled HHL circuits in Fig.~2(a) consist of more than 15 one- and two-qubit gates, excluding
phase compensation and tomography gates that are not shown.
Execution of these gates with reasonably high precisions are therefore important,
though phase errors may not be critical at certain steps, e.g., the process of mapping eigenvalues to the register.
Our single-qubit rotations have been calibrated by randomized benchmarking
to be of reasonably high fidelity as limited by qubit coherence~\cite{Xu2016}.
$Q_3$'s $\pi$ gate is 300 ns in length, and qubit coherence would limit the gate fidelity to be just under 0.98.
For the $Q_2$-$Q_3$ $\sqrt{\textrm{iSWAP}}$ and iSWAP gates,
we tune the qubits on-resonance
for periods of $\pi/\left(4g_{2,3}\right)$ and $\pi/\left(2g_{2,3}\right)$, respectively,
similar to those demonstrated using the qubit-resonator architecture~\cite{Hofheinz2009}.
However, here due to the crosstalk issue, which likely results from
insufficient crossover wirings to tie the grounds together~\cite{ChenZ2014},
our $\sqrt{\textrm{iSWAP}}$ and iSWAP gates show slightly lower performances,
with fidelity values estimated to be slightly above 0.98 and 0.97, respectively.
For the CZ gate we include the qubit's second excited state $\ket{2}$
and implement similarly as that demonstrated using the qubit-resonator-qubit architecture~\cite{Mariantoni2011}.
The fidelity of our CZ gate is estimated to be about 0.95 using the calibrated Ramsey interference data shown in Fig.~2(c).
We note that the CZ gate can also be implemented, with high fidelity
and without exchange of excitations, by optimizing one qubit's
frequency trajectory to mediate the two-qubit $\ket{11}$ state close to the avoided-level crossing
with the $\ket{02}$ state~\cite{Barends2014}, which would be hard to be implemented in our device due to the crosstalk issue.

We prepare 18 different input states $\ket{b_j}$, for $j = 1$ to 18, on the Bloch sphere as shown in Fig.~3(a).
With the output state stored in $Q_1$, we can measure the expectation value of
a certain operator that we are interested in (see, e.g., the case of $\ket{b_1} = \ket{0}$ in Fig.~3(b)).
We can also characterize the input and output states of the above instance by quantum state tomography (QST):
qubit polarization along Z axis of the Bloch sphere can be directly measured; for polarizations along
X and Y axes the qubit is rotated around Y and X axes, respectively, before measurement.
The corresponding output $\ket{x_j}$ in $Q_1$ and the state of $Q_4$ are simultaneously measured \cite{Supplement},
which allows us to reconstruct $\ket{x_j}$ using only the data corresponding to the $\ket{1}$-state outcome of $Q_4$.
As shown in Fig.~3(c), fidelity values of the 18 output states by QST are reasonably high, ranging from $0.840\pm0.006$ to $0.923\pm0.008$.

With all the input and output states being measured, we are able to
characterize the solver more exactly by QPT,
where a non-trace-preserving process matrix in the Pauli basis is used to linearly map $\ket{b_j}$ to $\ket{x_j}$.
The experimental $\chi_\textrm{exp}$ matrix, inferred reliably only with all 18 input states being used,
and the ideal matrix $\chi_\textrm{id}$ are shown together in Fig.~4 for comparison, which yields
a process fidelity~\cite{PhysRevLett.95.210505} of
$F = \textrm{Tr}\left(\chi_\textrm{id}\chi_\textrm{exp}\right)/\left[\textrm{Tr}\left(\chi_\textrm{id}\right)\textrm{Tr}\left(\chi_\textrm{exp}\right)\right] = 0.837 \pm 0.006$.
Here the error bar is estimated using simulated random normal distributed noise associated with the QST data~\cite{PhysRevLett.93.080502}.
$\chi_\textrm{exp}$ has a trace of 0.497, which indicates the averaged
success probability of our solver for the 18 input states, while ideally the trace is 0.625 for $C = 1$. \cite{Supplement}

\begin{figure}[t]
\includegraphics[width=0.7\linewidth]{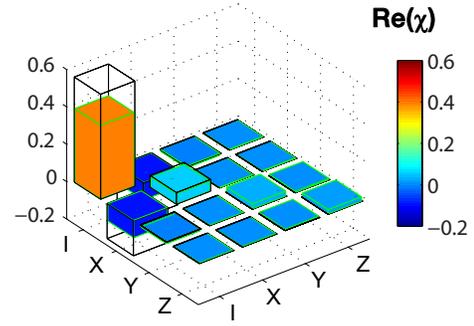}%
\caption{Non-trace-preserving QPT characterizing the quantum linear solver.
Shown are the real parts of the experimental $\chi_\textrm{exp}$ matrix (bars with color)
and the ideal $\chi_\textrm{id}$ matrix (black frames),
where $I$ is the identity and \{$X$, $Y$, $Z$\} are the Pauli operators \{$\sigma_x$, $\sigma_y$, $\sigma_z$\}.
All imaginary components (data not shown) of $\chi_\textrm{exp}$
are measured to be no higher than 0.015 in magnitude.
\label{figure_QPT}}
\end{figure}

In summary, we have demonstrated a superconducting quantum linear solver
for a $2\times2$ system with output state fidelities ranging
from $0.840\pm0.006$ to $0.923\pm0.008$ and quantum process fidelity of $0.837\pm0.006$.
The achieved fidelities are limited by decoherence and by errors in our implementation of the two-qubit gates,
the latter of which is related to the insufficient grounding that can be fixed by adding an extra
lithography layer in the device fabrication~\cite{Barends2014,Kelly2015}.
With future improvements on superconducting qubit coherence
and circuit complexity, the superconducting quantum circuits could be used
to implement more intricate quantum algorithms on a larger scale
and ultimately reach quantum computational speedup.

\textbf{Acknowledgments.}
We thank Masoud Mohseni for helpful discussions.
%Devices were made at the Nanofabrication Facilities at Institute of Physics in Beijing, University of Science and Technology in Hefei, and National Center for Nanoscience and Technology  in Beijing. Measurements were performed mainly at Zhejiang University in Hangzhou, with part of the test done at University of Science and Technology in Shanghai.
This work was supported by the Chinese Academy of Sciences, the NBRP of China (Grants No. 2014CB921201 and No. 2014CB921401),
the NSFC (Grants No. 11434008, No. 11374344, No. 11404386 and No. 11574380), and the Fundamental Research Funds for the Central Universities of China (Grant No. 2016XZZX002-01).

\end{document}